\documentstyle[12pt,fleqn]{article}
\begin{document}

\vspace*{0.3cm}
\begin{center}
  {\bf  RELATIVISTIC INVARIANT LIE ALGEBRAS
 FOR KINEMATICAL OBSERVABLES IN QUANTUM SPACE-TIME}
 
\bigskip

{\large
V.~V.~Khruschev$^{a,}$}
\footnote{e-mail: khru@uipe-ras.scgis.ru

\hspace{4pt}$^2$e-mail: andrey@buzon.uaem.mx}
{\large and A.~N.~Leznov$^{b,c,d,2}$}

\bigskip

 $^a$ {\it Department for Gravitation Researches in Metrology
VNIIMS,  3-1\\ M.Ulyanovoy Street, Moscow, 117 313,
Russia} 

$^b$ {\it Universidad Autonoma del Estado de Morelos,
CCICAp, Cuernavaca, Mexico}

$^c$ {\it Institute for High Energy Physics,  Protvino,
Moscow Region, 142280, Russia}

$^d$ {\it Bogoliubov Laboratory of Theoretical Physics, JINR,
Dubna, \\ Moscow Region, 141980, Russia}
\end{center}

\bigskip
\begin{abstract}
\noindent A deformation of the canonical algebra for kinematical
observables of the quantum field theory in Minkowski space-time has been 
considered under the
condition of Lorentz invariance. A  relativistic invariant algebra obtained
depends on  additional fundamental constants $M$, $L$ and  $H$ with the 
dimensions of mass, length and action, respectively. In some limiting cases 
the algebra goes over into the well-known Snyder or Yang algebras. In 
general case the algebra represents a class of Lie algebras, that consists
of simple algebras and semidirect sums of simple algebras and integrable ones. 
Some algebras belonging to this class are noninvariant under $T$ and 
$C$ transformations. Possible applications of obtained algebras for 
descriptions of states of matter under extreme conditions are
briefly discussed.
\smallskip

\noindent {\it PACS:} 11.30.Cp; 11.30.Er; 11.10.Cd; 02.20.Sv

\noindent {\it MSC(2000):} 81R05; 22E70
\smallskip

\noindent {\it Keywords:} Space-time symmetries; Fundamental constants; 
Lorentz invariance; Quantum space-time; Time reversal; Charge conjugation 
\end{abstract}
\medskip

\medskip

At present the concept of continuous Minkowski space-time together
with the group of its motions, namely the Poincar{\'{e}} group, is basic for
description of all established physical phenomena in the framework of 
relativistic theory.
The space-time properties of any physical object according to this concept
are rather simple and well understood. But it is not so, when one would
proceed to the investigation of unexplored states of matter under extreme 
conditions,
for instance, in early universe or in quark-gluon plasma [1, 2]. The
space-time properties of these systems may be more complicated and 
probably determinated
by a generalized group of space-time symmetries in extra dimensions [3].
Moreover their space-time properties may depend on additional fundamental
constants as compared with the light velocity $c$ and Planck constant of
action $\hbar$. 

In this paper we present the results of study of possible generalizations
of conventional space-time symmerties. In order to minimize a number of
generalized symmetries we take into account 
some restrictions such as a preservation, if it is possible, of all 
known physical properties of observables and a fulfilment of the matematical  
condition that  observables should generate a Lie group.
The Poincar{\'{e}} group is a semidirect product of the Lorentz group and
the Abelian four-dimensional translation group in Minkowski space-time. In
quantum case the Heisenberg commutation relations allow one to consider 
coordinates as physical operators and on the same footing as  momenta.
Moreover, it is known  that  elaboration of the measurement procedure for the
observables within an atom made Born to formulate the reciprocity principle
for  coordinates and  momenta [4]. In the framework of the canonical
quantum relativistic theory all fundamental space-time properties can be
represented by a system of commutation relations between the Hermitian
operators of the following 15-dimensional algebra: 
\[
\hspace{5em}[x_{i},x_{j}]=[p_{i},p_{j}]=0,\quad \lbrack p_{i},x_{j}]=i\hbar
g_{ij}I,
\]%
\[
\hspace{5em}[p_{i},I]=[x_{i},I]=[F_{ij},I]=0,
\]%
\begin{equation}
\hspace{5em}[F_{ij},F_{kl}]=i\hbar
(g_{jk}F_{il}-g_{ik}F_{jl}+g_{il}F_{jk}-g_{jl}F_{ik}),
\end{equation}%
\[
\hspace{5em}[F_{ij},p_{k}]=i\hbar (g_{jk}p_{i}-g_{ik}p_{j}),
\]%
\[
\hspace{5em}[F_{ij},x_{k}]=i\hbar (g_{jk}x_{i}-g_{ik}x_{j}),
\]%
\noindent where $x_{i}$ and $p_{i}$ are the operators of  4-coordinates and
4-momenta, respectively, $F_{ij}$ are the proper Lorentz group generators, 
$I$ is the ''identity'' operator.

The  space-time points may be discriminated as  eigenvalues 
of the $x_{i}$ generators for certain irreducible representation of the 
algebra (1). For instance, 
in the $x$ - representation the basis vectors of the representation have
the form $\psi _{\alpha \beta }(x)$, where $x=\{x_{0},x_{1},x_{2},x_{3}\}$
are the eigenvalues of operators for 4-coordinates, $\alpha ,\beta $ are the
discrete spin indices, which are subject to an action of $S_{ij}$, where $%
S_{ij}=F_{ij}-x_{i}p_{j}+p_{i}x_{j}$ are the spin operators of some finite
dimensional representation of the Lorentz group.

The conventional quantum theory, which is based on the system of 
commutation relations (1), has a long standing problem concerning
a removal of shot-distance singularities. It was the reason, which induced
Heisenberg to suggest the idea that the configuration-space coordinates
may not commute [5].   
Then Snyder  introduced the Lorentz invariant quantized 
space-time characterized by a fundamental length [6].
 This theory is, however, not invariant under  translations. 
 Yang presented a generalized translation invariance 
and suggested a new fundamental unit of mass [7]. In the limiting
case, when the fundamental length and mass are eliminated from the theory,
the standard commutation relations (1) are valid between the kinematical 
observables.
But the Snyder and Yang theories are not  maximal generalizations of the quantum
relativistic theory providing a strict Lorentz invariance. In the
work [8] a more general algebra for the kinematical observables has been
found and a new fundamental unit of action has been defined.

Below we clear up the presumptions used in
a deduction of  deformed relativistic
invariant Lie algebras for  observables of  quantum theories,
generalize to a certain extent the results obtained in Ref. [8], 
and point to some possible
applications of the generalized algebras. The  obtained 
class of Lie algebras consists of  simple algebras as well as 
semidirect sums of simple algebras and integrable ones. When a new 
constant with the dimensions of action enter in  generalized commutation 
relations, the corresponding algebras became noninvariant under the time 
reversal $T$ or the charge conjugation $C$.

We suppose that a generalization of the algebra (1) is performed under
the following conditions:

1. The generalized algebra should be a Lie algebra.

2. The dimensionality of the algebra, which is subject to a generalization,
and the physical dimensions of  operators entered in it should be
preserved.

3. The generalized algebra should contain the Lorentz algebra as its
subalgebra, and  commutation relations of the Lorentz algebra with other
generators should be the same as for the initial algebra.

The procedure of the generalization of the algebra (1) described above
may be named a
relativistic or Lorentz invariant deformation of the algebra (1) because of the
property of Lorentz symmetry is conserved as a fundamental law of nature.
Note that in some cases the canonical Poincar{\'{e}} invariance may be violated.

Under these conditions algebra spanned on
the Hermitian operators $F_{ij}$, $p_{i}$, $x_{i}$ and $I$, which is
the maximal generalization of algebra (1), has the following form:

\[
\hspace{+5em} [F_{ij},F_{kl}] = \varphi
(g_{jk}F_{il}-g_{ik}F_{jl}+g_{il}F_{jk}-g_{jl}F_{ik}), 
\]
\[
\hspace{+5em} [p_i,x_j] = Ag_{ij}I+BF_{ij}+C\epsilon_{ijkl}F_{kl}, 
\]
\[
\hspace{+5em} [p_i,p_j] = aF_{ij}+b\epsilon_{ijkl}F_{kl}, 
\]
\[
\hspace{+5em} [x_i,x_j] = cF_{ij}+d\epsilon_{ijkl}F_{kl}, 
\]
\begin{equation}
\hspace{+5em} [p_i,I] = \alpha x_i+\beta p_i,  \label{medalg}
\end{equation}
\[
\hspace{+5em} [x_i,I] = \gamma x_i+\delta p_i, 
\]
\[
\hspace{+5em} [F_{ij},x_k] = h(g_{jk}x_i - g_{ik}x_j), 
\]
\[
\hspace{+5em} [F_{ij},p_k] = f(g_{jk}p_i - g_{ik}p_j), 
\]
\[
\hspace{+5em} [F_{ij},I] = 0, 
\]

\noindent where $\epsilon_{ijkl}$ is the Levi-Civita tensor.

The relations (\ref{medalg}) contain fourteen arbitrary pure imaginary
parameters. Taking into account the Jacobi identities and the dimensions of
physical operators entered in the commutation relations (\ref{medalg}), 
ten parameters must be excluded, and the following relativistic invariant
algebra, that is a maximal deformation of algebra (1) under conditions 1-3, 
can be obtained: 
\[
\hspace{5em}%
[F_{ij},F_{kl}]=if(g_{jk}F_{il}-g_{ik}F_{jl}+g_{il}F_{jk}-g_{jl}F_{ik}),
\]%
\[
\hspace{5em}[p_{i},x_{j}]=if(g_{ij}I+\frac{F_{ij}}{H}),
\]%
\[
\hspace{5em}[p_{i},p_{j}]=\frac{if}{L^{2}}F_{ij},
\]%
\[
\hspace{5em}[x_{i},x_{j}]=\frac{if}{M^{2}}F_{ij},
\]%
\begin{equation}
\hspace{5em}[p_{i},I]=if(\frac{x_{i}}{L^{2}}-\frac{p_{i}}{H}),
\label{finalg}
\end{equation}%
\[
\hspace{5em}[x_{i},I]=if(\frac{x_{i}}{H}-\frac{p_{i}}{M^{2}}),
\]%
\[
\hspace{5em}[F_{ij},p_{k}]=if(g_{jk}p_{i}-g_{ik}p_{j}),
\]%
\[
\hspace{5em}[F_{ij},x_{k}]=if(g_{jk}x_{i}-g_{ik}x_{j}),
\]%
\[
\hspace{5em}[F_{ij},I]=0.
\]

The algebra (\ref{finalg}) depends on four dimensional parameters: $L$ is a
constant with the dimensions of length, $M$ is a constant with the
dimensions of mass, $H$ and $f$ are the constants with the dimensions of
action ($M$ and $L$ take  real values as well as pure imaginary ones, $c=1$ in
the system of units being used). 

In general case the algebra (\ref{finalg}) can be considered 
as the algebra of observables for the Lorentz invariant quantum theory with
noncommutative coordinates and momenta.
In the limiting case, when $M$, $L$ and $H$
values become infinitely large, the commutation relations (\ref{finalg}) go 
over into the commutation relations of contracted 
algebra (1) providing $f=\hbar $. 
A more complicated case is also possible, when $f$ is some
function versus three parameters $L$, $M$ and $H$. However, in order that an
agreement with the conventional commutation relations should take place in
the limiting case, as $L\rightarrow \infty $, $M\rightarrow \infty $ 
and $H\rightarrow \infty $, the value of $f(L,M,H)$ must be equal to $\hbar $.

In other limiting cases, when $f=\hbar $, but $M$, $L$ and $H$ have
different magnitudes, the following theories may be obtained:

\noindent a) $H\rightarrow \infty $, $L\rightarrow \infty $ - the relativistic
quantum theory with noncommutative coordinates;

\noindent b) $H\to \infty$, $M\to \infty$ - the relativistic quantum theory 
with noncommutative  momenta;

\noindent c) $H\rightarrow \infty $ - the relativistic quantum theory 
with noncommutative coordinates and momenta.

From mathematical point of view the system of commutation relations 
(\ref{finalg}) specify some class of Lie algebras, which consist of semisimple 
algebras as well as general type algebras. After the calculation of the
Killing-Cartan form the condition for semisimplicity may be written as

\begin{equation}
\hspace{10em}\frac{f^{2}(M^{2}L^{2}-H^{2})}{H^{2}M^{2}L^{2}}\neq 0.
\label{killing}
\end{equation}

\noindent If one performs the linear transformation 
of $p_{i},x_{i},I$  generators in the form:

\begin{equation}
\hspace{4em}F_{i5}=Bx_{i}+Dp_{i},\quad F_{i6}=Ex_{i}+Gp_{i},\quad F_{56}=AI,
\end{equation}

\noindent then under the condition (\ref{killing}) one may obtain the
commutation relations for the algebras of pseudoorthogonal 
groups $O(3,3)$, $O(2,4)$ and $O(1,5)$. 
These algebras correspond to the definite values of $M^{2},L^{2},H^{2}$ 
parameters, which are shown in Table 1.

\newpage
{\small Table 1. \quad{} }

{\small 
\parbox[t]{13.9cm}{\small The real simple Lie algebras that correspond to 
the different values of  $L^2$, $M^2$ \\and  $H^2$ parameters.}}

\begin{center}
\begin{tabular}{|l|c|}
\hline
\qquad \quad $M^{2}$,\quad $L^{2}$ and\quad $H^{2}$  values & Algebra\quad 
\\[5pt]\hline\hline
\quad $H^{2}<M^{2}L^{2}$, \quad $M^{2}>0$, \quad $L^{2}>0$ & \qquad $%
o(2,4)\qquad $ \\[5pt]\hline
\quad $H^{2}<M^{2}L^{2}$, \quad $M^{2}<0$, \quad $L^{2}<0$ & \qquad $%
o(2,4)\qquad $ \\[5pt]\hline
\quad $M^{2}>0$,\quad $L^{2}<0$, \quad or\quad $M^{2}<0$,\quad $L^{2}>0$
\quad  & \qquad $o(2,4)\qquad $ \\[5pt]\hline
\quad $H^{2}>M^{2}L^{2}$, \quad $M^{2}>0$, \quad $L^{2}>0$ & \qquad $%
o(1,5)\qquad $ \\[5pt]\hline
\quad $H^{2}>M^{2}L^{2}$, \quad $M^{2}<0$, \quad $L^{2}<0$ & \qquad $%
o(3,3)\qquad $ \\[5pt]\hline
\end{tabular}
\end{center}

\smallskip

For $H^{2}=M^{2}L^{2}$ and $M^{2}>0$, $L^{2}>0$ the $o(1,5)$ algebra
degenerates into a semidirect product of the $o(1,4)$ algebra and the algebra
of 5 - translations, while for $H^{2}=M^{2}L^{2}$ 
and $M^{2}<0$, $L^{2}<0$ the $o(3,3)$ algebra degenerates into a semidirect 
product of the $o(2,3)$ algebra and the algebra of 5 - translations. 
Note that a transition to the limit $A_{\alpha }\rightarrow \infty $, 
where $A_{\alpha }$ is any 
term of the set $\{M^{2},$ $L^{2},$ $H^{2},$ $(M^{2},L^{2})\}$, 
do not exclude the algebras (\ref{finalg}) from the class of simple algebras, 
as distinct from the transitions $B_{\alpha }\rightarrow \infty $, 
where $B_{\alpha }\in $ 
$\{(M^{2},H^{2}),$ $(L^{2},H^{2}),$ $(M^{2},L^{2},H^{2})\}$ 
or $M^{2}L^{2}\rightarrow H^{2}$.

The irreducible representations of algebras (\ref{finalg}) are determined
with the help of  eigenvalues of Casimir operators. For the real
simple algebras shown in Table 1 the Casimir operators
have the known forms in terms of the generators $F_{ij}$, $i,j=0,1,...,5 $ 
of  pseudoorthogonal groups in six-dimensional spaces: 

\begin{equation}
\hspace{3em} K_1=\epsilon_{ijklmn}F^{ij}F^{kl}F^{mn}, \quad K_2=F_{ij}F^{ij},
\quad K_3=(\epsilon_{ijklmn}F^{kl}F^{mn})^2.
\label{cas}
\end{equation}

These operators  can also be expressed through  $p_i, x_i, F_{ij}$,
$i,j=0,...,3$, and $I$ generators. For instance, the second-order
invariant operator may be represented in the form:

\begin{equation}
\hspace{3em} C_2=\sum_{i<j}F_{ij}F^{ij}(\frac{1}{M^{2}L^{2}}-
\frac{1}{H^{2}})+I^{2}+\frac{x_{i}p^{i}+p_{i}x^{i}}{H}-
\frac{x_{i}x^{i}}{L^{2}}-\frac{p_{i}p^{i}}{M^{2}},
\label{casim}
\end{equation}

\noindent that in the limit case $M\rightarrow \infty $, 
$L\rightarrow \infty ,$ $H\rightarrow \infty $ go over into the canonical
 ''identity'' operator squared (1). The limit expressions of
Casimir operators (\ref{cas}), 
as $L\rightarrow \infty $, $M\rightarrow \infty $, 
have been found in Ref. [9]  in the case $H =
\infty $, while the explicit form for the quadratic Casimir operator 
(\ref{casim}) in this
case has been presented in Ref. [10]. The quadratic Casimir operator for
an algebra  in three space dimensions, which is a particular case of the
algebra (\ref{finalg}), has been obtained in Ref. [11].

It should be noted that the presence of a new constant $H$ with
the dimensions of action leads to the noninvariance of  
system (\ref{finalg}) under the $T$ and $C$ 
transformations [5]. Indeed, the time
reversal results in  sign changes for all dimensional quantities, which
include the time variable in an odd degree. Evidently, if simultaneously with
the time reversal the appropriate transformations corresponding to this
reversal for the physical operators have been done, then the system of
commutation relations will remain invariant. In the conventional theory 
behaviour of the commutation relations (1) with respect to the time reversal
is determined by the Planck constant, so the $T$ transformation is
equivalent to the sign change for $\hbar $. However, the transition to 
conjugate or transposed operators simultaneously with the time reversal
preserves the commutation relations (1).

For the algebra (\ref{finalg}) there is the additional 
 constant $H$, which is also
odd with respect to the time reversal and enters in (\ref{finalg})  so
 that it is impossible to restore the $T$ invariance of the system 
(\ref{finalg}) for $H\neq \infty $. 
Along the same lines one may obtain the $C$
noninvariance of the system (\ref{finalg}), since the quantities with
dimensionality of mass change its signs after  replacement of particles
by antiparticles. $TC$ transformation do not change $\hbar $ and $H$, thus
the system (\ref{finalg}) is invariant under $TC$ and $P$
transformations.

Moreover the generalization of the algebra (1) with $L\neq \infty $ leads to a
breakdown of Poincar{\'{e}} invariance, as follows from the commutation 
relations (\ref{finalg}).
The investigation of the quantum theory in momentum space with constant
curvature (which is related to the algebra (\ref{finalg}) as $H\rightarrow
\infty $, $M\rightarrow \infty $), including  modifications to Feynman
rules and the locality principle,  has  been started 
in Refs. [12, 13]. A connection of thermal properties of quantum system 
with the generalized space-time geometry was  investigated as well 
[see, e.g., 14]. 
If the values of the parameters entering in the system (\ref{finalg}) 
are of  cosmic scales under the condition that the usual physical
phenomena take place at least at distances of the order of Solar system, then,
for instance, the value of  $M$ parameter should be of the order of 
 the mass of the Universe and
 the associated elementary length $\hbar /M$ 
 turns out much less than typical nuclear distances and time
intervals. 
Additional constants with the dimensions of length 
$H/M$ and $\hbar L/H$ or
of mass $H/L$ and $\hbar M/H$  may also have some meaning. Their numerical
values by no means are limited by the correspondence principle with the
conventional theory for macroscopical phenomena. Moreover, one may 
 consider  algebras of type (\ref{finalg}) for the description of such
objects as quarks or other colour particles, which never have been observed in
the conventional space-time in free states. Therefore some stringent
restrictions for violation of the standard quantum theory principles, which
take place at present for  usual elementary particles, cannot be applied.
In this case the values of parameters $\hbar /L$ and $\hbar /M$ might be of
the order of 1~GeV and 1~Fm respectively [10, 15]. It is known, that at present
 there is a
possibility, that the problem of confinement of the colour objects cannot be
resolved in the framework of quantum chromodynamics (QCD)  itself and, 
perhaps, for its resolution some extra postulates are needed. 
The use of algebra (\ref{finalg}) 
instead of algebra (1) makes it possible to transform the problem
of confinement, which is a dynamical problem in the framework of QCD, to a
kinematical one. Moreover, one can to generalize the class of deformed
algebras for quarks or elementary particles  by means of weakening of the 
conditions
under which the algebra (\ref{finalg}) was found. For instance,
in Refs. [16, 17] the number of generators of generalized algebra
has been enlarged and the semidirect products of $su(1,3)$ algebra
and some integrable algebras, that contains the operators of coordinates
and momenta, have been considered. If one preserve the number of 
generators, the $su(1,3)$ algebra by itself is the generalization of the
 commutation relations (1), which is performed by weakening of
the third condition just to the requirement
that the generalized algebra should contain the Lorentz subalgebra.

The Coulomb  and  harmonic oscillator problems
were considered in the work [11] 
in the three-dimensional quantum space described by $O(5)$ algebra.
It has been shown that the energy spectrum of the Coulomb problem with
conserving the Runge-Lenz vector coincides with a part of the spectrum found
by Schr{\"{o}}dinger for the space with a constant curvature. 
Some methods for evaluating  matrix elements of operators  in quantum
space with the dimensional parameters $L$, $M$ and $H$ were worked out as well.
In the last years a number of results were  obtained 
in the framework of approaches    
with  noncommutative configuration-space variables
({\it e.g.}, [18-22]), which have been initiated by  string theory,
noncommutative geometry and quantum gravitation and operate, as a rule, with
more general algebraic structures than  Lie algebras.

\bigskip
\noindent{\large\bf References}

\vspace{0.5cm}

\noindent [1]  J. Silk, The Big Bang,  3rd ed., W.H. Friman \& Co., 
 2001.

\noindent [2]  Proc. of Quark Matter Conf.,  18-24 July, Nantes, France,
 2002.

\noindent [3]  V.A. Rubakov and  M.E. Shaposhnikov,
 Phys. Lett.  B125, 136 (1983).

\noindent [4]  M.  Born, Proc. Roy. Soc. A165, 291 (1938).

\noindent [5]  R. Jackiw, physics/0209108.

\noindent [6]  H. Snyder, Phys. Rev. 71, 38  (1947).

\noindent [7]  C.N. Yang, Phys. Rev. 72, 874 (1947).

\noindent [8]  A.N. Leznov and  V.V. Khruschev, 
 {\it preprint} IHEP 73-38, Serpukhov, 1973. 

\noindent [9]  A.N. Leznov, JETP Lett. 45, 321 (1965).

\noindent [10]  V.V. Khruschev, Grav. \& Cosmol. 3, 197 (1997).

\noindent [11]  A.N. Leznov, Nucl. Phys. B640, 469 (2002).

\noindent [12]  Yu.A. Golfand, JETP. 37, 504 (1959).

\noindent [13]  V.G. Kadyshevsky, JETP. 41, 1885 (1961); 
{\it in}: {I.~E.~Tamm} memorial vol.

 "Problems of Theoretical Physics", eds, V.I. Ritus, E.L. Feinberg, 

V.L. Ginsburg et al. Nauka, M. 1972.

\noindent [14]  B.S. Kay and  R.M. Wald, Phys. Rep. 207, 49 (1991).

\noindent [15]  V.V. Khruschev,  V.I. Savrin and  S.V. Semenov, 
Phys. Lett. B525, 283 (2002).

\noindent [16]  V.V. Khruschev,  J. Nucl. Phys.  46, 219 (1987).

\noindent [17] S.G. Low,  J. Phys. A35, 5711 (2002).


\noindent [18]  N. Seiberg and  E. Witten, J. High Ener. Phys.
 09-032, 1 (1999).

\noindent [19]  R. Oeckl, Nucl. Phys. B581, 559 (2000).

\noindent [20]  P. Kosi\'nski,  J. Lukierski and  P. Ma\'slanka, hep-th/0012056.

\noindent [21]  S. Doplicher,  K. Fredenhagen and  J. Roberts,
Phys. Lett.  B331, 39 (1994).

\noindent [22]  L.J. Garay, Int. J. Mod. Phys. A10, 145  (1995).

\end{document}